\documentclass[aps,prd,superscriptaddress,showpacs,showkeys]{revtex4} 
\usepackage{latexsym,epsfig,amsmath,amssymb} 
\usepackage{amsfonts} 
\usepackage{verbatim} 
 
\begin{document} 
 
\title{Parameters of scalar resonances from the combined analysis 
of data on processes $\pi\pi\to\pi\pi,K\overline{K},\eta\eta$ and 
$J/\psi$~decays} 
 
\author{Yurii S. Surovtsev} 
\affiliation{Bogoliubov Laboratory of Theoretical Physics, Joint 
Institute for Nuclear Research, 141 980 Dubna, Russia} 
\author{Petr Byd\v{z}ovsk\'y} 
\affiliation{Nuclear Physics Institute, Czech Academy of Sciences, 25068 
\v{R}e\v{z}, Czech Republic} 
\author{Robert Kami\'nski} 
\affiliation{Institute of Nuclear Physics, Polish Academy of Sciences, 
Cracow 31342, Poland} 
\author{Valery E. Lyubovitskij} 
\affiliation{Institut f\"ur Theoretische Physik, 
Universit\"at T\"ubingen, 
Kepler Center for Astro and Particle Physics,\\ 
Auf der Morgenstelle 14, D-72076, T\"ubingen, Germany} 
\affiliation{Department of Physics, Tomsk State University, 
634050 Tomsk, Russia} 
\author{Miroslav Nagy} 
\affiliation{Institute of Physics, Slovak Academy of Sciences, Bratislava 
 84511, Slovak Republic} 
 
\date{\today} 
 
\begin{abstract} 
A combined analysis of data on the isoscalar $S$-wave processes 
 $\pi\pi\to\pi\pi,K\overline{K},\eta\eta$ and on decays 
 $J/\psi\to\phi\pi\pi,\phi K\overline{K}$ from the DM2, Mark III 
 and BES II Collaborations is performed to study $f_0$ mesons. 
 The method of analysis is based on analyticity and unitarity and 
 uses an uniformization procedure. In the analysis limited only 
 to the multichannel $\pi\pi$-scattering data, two possible sets 
 of parameters of the $f_0(500)$ were found: In both cases the mass 
 was about 700~MeV but the total width was either about 680 or 1040~MeV. 
 The extension of the analysis using only the DM2 and Mark III data on 
 the $J/\psi$ decays does not allow us to choose between these sets. 
 However, the  data from BES II on the di-pion mass distribution 
 in the decay $J/\psi\to\phi\pi^+\pi^-$ clearly prefer the wider 
 $f_0(500)$ state. Spectroscopic implications from the results of 
 the analysis are also discussed. 
 
\end{abstract} 
 
 
 
\maketitle 
 
\section{Introduction} 
 
A comprehension of the nature of the scalar mesons is very important for 
some major topics in particle physics such as the QCD vacuum. However, 
both parameters of the scalar mesons, obtained from experimental data 
in various analyses, and even the status of some of them, are 
still quite ambiguous~\cite{PDG-12,SBL-prd12}. 
As for the meson parameters, let us mention the widely discussed 
$f_0(500)/\sigma$ meson (formerly $f_0(600)$), $f_0(980)$, and 
$f_0(1500)$. A doubtful meson existence can be demonstrated on 
the case of the $f_0(1370)$ state which is apparently required 
by a bulk of data~\cite{Bugg07}, but in some analyses of only 
the $\pi\pi$ scattering no evidence for its existence was 
found~\cite{Ochs_Mink}. We have shown that the existence of 
the $f_0(1370)$ does not contradict the data on  
$\pi\pi\to\pi\pi\,,K\overline{K},\eta\eta,\eta\eta^{\prime}$~\cite{SBL-prd12}. 
In the hidden gauge unitary approach, 
the $f_0(1370)$ appears dynamically generated as a $\rho\rho$ 
state~\cite{Geng_Oset} and the $f_0(1710)$ as generated from the 
$K^\ast \bar K^\ast$ interaction. 
 
Note also a situation with scalar states in the 1500-MeV region. 
In our previous model-independent analyses of 
$\pi\pi\to\pi\pi,\,K\overline{K},\eta\eta(\eta\eta^{\prime})$, 
we observed a wide state $f_0(1500)$ whereas in some other analyses, 
which included mainly meson production and decay processes and which 
are cited in the PDG tables \cite{PDG-12}, a rather narrow $f_0(1500)$ 
was obtained. We have suggested that the wide $f_0(1500)$, observed in 
the multichannel $\pi\pi$ scattering, is effectively a superposition 
of two states, the wide and narrow state. The latter is observed only 
in decays and productions of mesons. 
This suggestion was verified in the model-independent two-channel 
analysis of data on $\pi\pi\to\pi\pi,K\overline{K}$~\cite{SBKLN_1206_3438}. 
In the presented article we confirm this assumption in the three-channel 
analysis of data on $\pi\pi\to\pi\pi,K\overline{K},\eta\eta$ and decays 
$J/\psi\to\phi\pi\pi,\phi K\overline{K}$ from the DM2, Mark III, and 
BES II Collaborations~\cite{MarkIII,DM2,BES}. This is necessary, 
especially, as the wide states provoke many questions 
which should be answered. 
 
In view of this situation related to parameters and the status 
of the scalar mesons, there are still many unsolved problems 
as to determining a QCD nature of the mesons and their assignment 
to the quark-model configurations in spite of a big amount of work 
devoted to these problems (see, e.g., \cite{Anisovich06} and references 
therein). 
 
In this article, we describe the multichannel $\pi\pi$ scattering  
($~\pi\pi\to\pi\pi,K\overline{K},\eta\eta~$) using the method based 
only on the first principles, analyticity and unitarity~\cite{KMS-96}, 
which allows us to avoid any theoretical prejudice in extracting the 
resonance parameters. This we call 
``the model-independence''~\cite{SBL-prd12,SBKLN_1206_3438,SBGKLN-1311.1066}. 
The method is applied to the analysis of experimental data on 
the multichannel $\pi\pi$ scattering and decays 
$J/\psi\to\phi\pi\pi,\phi K\overline{K}$. 
The $J/\psi$ decays are described using a formalism from 
Refs.~\cite{MP-prd93,Zou-Bugg-prd94}, where certain reasonable 
assumptions about the final-state interactions are made.  
Considering the obtained arrangement of resonance poles on the 
Riemann-surface sheets, the constants of resonance couplings 
with the channels, and the resonance masses we can draw definite conclusions 
about the nature of the investigated states. 
 
The article is organized as follows. A basic formalism for the 
three-channel model-independent method was already given in our 
previous paper \cite{SBL-prd12}; therefore, in Sec.~II we give 
only formulas introducing parameters determined in the analysis. 
Results of the combined coupled-channel analysis of data on isoscalar 
$S$-wave processes $\pi\pi\to\pi\pi,K\overline{K},\eta\eta$ and on decays 
$J/\psi\to\phi\pi\pi,\phi K\overline{K}$ are presented in Sec.~III. 
Discussion of the results and conclusions are given in Sec.~IV. 
 
\section{The three-coupled-channel formalism in model-independent 
approach with uniformizing variable} 
 
The multichannel $S$-matrix can be described in our model-independent 
method, which essentially utilizes a uniformizing variable, without any 
approximations only in the two-channel case. 
In the three-channel case, a four-sheeted model of the eight-sheeted Riemann 
surface has to be constructed to obtain a simple symmetric (easily interpreted) 
picture of the resonance poles and zeros of the $S$ matrix on 
the uniformization plane. 
The matrix elements $S_{ij}$, where $i,j=1,2,3$ denote the channel numbers, 
have the right-hand cuts along the real axis of the $s$ complex plane 
($s$ is the invariant total energy squared) starting with the channel 
thresholds $s_i$ and the left-hand cuts related to crossed channels. 
An influence of the lowest branch point $s_1$ ($\pi\pi$) is neglected 
but unitarity on the $\pi\pi$ cut is kept. 
Sheets of the Riemann surface are numbered according to the signs of 
analytic continuations of the square roots $\sqrt{s-s_i}~~\mbox{as follows:}$ 
$ \mbox{signs}\Big(\mbox{Im}\sqrt{s-s_1}, 
\mbox{Im}\sqrt{s-s_2},{\mbox{Im}}\sqrt{s-s_3}\Big)= 
+++,-++,--+,+-+,+--$, $---,-+-, ++- $~ correspond to sheets 
I, II,$\cdots$, VIII, respectively. 
 
Resonances are described on the Riemann surface using the formulas for 
analytic continuations of the $S$-matrix elements to all sheets. 
The formulas allow us to express the matrix elements on the unphysical 
sheets by means of the matrix elements on the physical sheet that 
have only the resonance zeros (aside the real axis), at least, around 
the physical region~\cite{SBL-prd12,KMS-96}. 
Assuming the resonance zeros on sheet I, we can obtain an arrangement 
of poles and zeros of the resonance  on the whole Riemann surface which we 
denote as a resonance cluster. In the three-channel case, we obtain 
{\it seven types} of the resonance clusters corresponding to possible 
situations when there are resonance zeros on sheet I only in $S_{11}$ -- ({\bf a}); 
~$S_{22}$~--~({\bf b});~$S_{33}$~--~({\bf c}); ~$S_{11}$ and $S_{22}$~--~({\bf d}); 
~$S_{22}$ and $S_{33}$~--~({\bf e}); ~$S_{11}$ and $S_{33}$~--~({\bf f}); 
~$S_{11}$, $S_{22}$, and $S_{33}$~--~({\bf g}). 
A three-channel resonance has to be described by one of the seven types of 
the resonance clusters which is the necessary and sufficient condition 
for its existence. 
The resonances of types ({\bf a}), ({\bf b}) and ({\bf c}) can be related 
to the resonances represented by Breit-Wigner forms but the types ({\bf d}), 
({\bf e}), ({\bf f}) and ({\bf g}) do not have their equivalents in the 
Breit-Wigner description. 
 
The cluster type is related to the nature of state. Considering the 
$\pi\pi$, $K\overline{K}$ and $\eta\eta$ channels, e.g., a resonance 
coupled relatively more strongly to the $\pi\pi$ channel than to the 
$K\overline{K}$ and $\eta\eta$ channels is described by the cluster 
of type ({\bf a}) but in the opposite case, the resonance is 
represented by the cluster of type ({\bf e}) (e.g., the state with 
the dominant $s{\bar s}$ component). The glueball must be represented 
by the cluster of type ({\bf g}) as a necessary condition for the ideal 
case. 
 
It is also possible to distinguish, in a model-independent 
way~\cite{KMS-96,MP-prd93}, a bound state of colourless particles 
({\it e.g.}, $K\overline{K}$ molecule) from a $q{\bar q}$ bound state. 
Alike in the one-channel case, 
the existence of the particle bound state means presence of a pole 
on the real axis below the threshold on the physical sheet. 
In the two-channel case, therefore, the existence of the bound state 
in channel 2 (e.g., $K\overline{K}$ molecule) that can decay into 
channel 1 ($\pi\pi$ decay) implies the presence of the pair of complex 
conjugate poles on sheet II below the second-channel threshold 
without the corresponding shifted pair of poles on sheet III. 
In the three-channel case, the bound state in channel 3 ($\eta\eta$) 
that can decay into the channels 1 ($\pi\pi$ decay) and 2 
($K\overline{K}$ decay) is represented by the pair of complex 
conjugate poles on sheet II and by the pair of shifted poles on 
sheet III below the $\eta\eta$ threshold without the corresponding 
poles on sheets VI and VII. 
 
The formulas of the analytic continuations \cite{SBL-prd12,KMS-96} 
prescribe that the resonance parameters (mass, total width, and coupling 
constants with the channels) must be calculated using the pole positions 
on sheets II, IV, and VIII because only on these sheets do the analytic 
continuations have the forms: $\propto 1/S_{11}^{\rm I}$, 
$\propto 1/S_{22}^{\rm I}$ and $\propto 1/S_{33}^{\rm I}$, respectively, i.e., 
the positions of poles on these sheets are at the same points 
of the complex-energy plane as the resonance zeros on the physical sheet. 
The other pole positions are shifted due to the coupling of channels. 
 
The $S$-matrix elements of all coupled processes are expressed 
in terms of the Jost matrix determinant $d(\sqrt{s-s_1},\cdots,\sqrt{s-s_n})$ 
using the Le Couteur-Newton relations \cite{LeCou}. 
The Jost determinant is a real analytic function with the only 
square-root branch points at $\sqrt{s-s_i}=0$. 
The important branch points, which correspond to the thresholds 
of the coupled and crossed channels, are taken into account in 
the uniformizing variable. In the uniformizing 
variable used here we neglect the lowest $\pi\pi$-threshold 
branch point but take into account the threshold branch points  
related to the two remaining channels, and the left-hand branch point 
at $s=0$ \cite{SBL-prd12} 
\begin{equation}\label{w} 
w=\frac{\sqrt{(s-s_2)s_3} + \sqrt{(s-s_3)s_2}}{\sqrt{s(s_3-s_2)}}, 
\end{equation} 
here $s_2=4m_K^2$, ~~$s_3=4m_\eta^2$. 
This variable maps our model of the eight-sheeted Riemann surface onto 
the uniformization $w$ plane divided into two parts by a unit circle centered 
at the origin. The semisheets I (III), II (IV), V (VII) and VI (VIII) are 
mapped onto the exterior (interior) of the unit disk in the 1st, 2nd, 3rd 
and 4th quadrants, respectively. The physical region extends from the $\pi\pi$ 
threshold 
$i(m_\eta\sqrt{m_K^2-m_\pi^2}+m_K\sqrt{m_\eta^2-m_\pi^2})/m_\pi\sqrt{m_\eta^2-m_K^2}$ 
on the imaginary axis along this axis down to the point {\it i} on the unit 
circle ($K\overline{K}$ threshold). Then it goes along the unit circle clockwise 
in the 1st quadrant to point 1 on the real axis ($\eta\eta$ threshold) 
and then along the real axis to the point 
$b=(\sqrt{m_\eta+m_K})/\sqrt{m_\eta-m_K}$ which is an image of $s=\infty$. 
The intervals ~$(-\infty,-b\,]$, $[-b^{-1}$, $b^{-1}]$, $[\,b,\infty)$ 
are the images of the corresponding edges of the left-hand cut of the 
$\pi\pi$-scattering amplitude. Each resonance is represented in $S_{11}$ 
by the poles and zeros that are symmetric to each other with respect to 
the imaginary axis. The representations of all possible types of resonances 
in $S_{11}$ on the $w$ plane can be found in Ref.~\cite{SBL-prd12}. 
 
The main model-independent effect of multichannel resonances is given 
by the pole clusters. Assuming that possible small remaining (model-dependent) 
contributions of resonances can be included via the background, 
the $S$-matrix elements are taken as the products 
\begin{equation} 
\label{factorisation} 
S=S_B S_{res} 
\end{equation} 
\label{d_res} 
where $S_B$ describes the background and $S_{res}$ the resonance contributions. 
 
On the $w$ plane, the Le Couteur-Newton relations are somewhat modified 
taking account of the used model of the initial eight-sheeted Riemann 
surface 
\begin{eqnarray} \label{LeCouteur-Newton} 
&&S_{11}=\frac{d^* (-w^*)}{d(w)},~~\qquad S_{22}=\frac{d(-w^{-1})}{d(w)}, 
~~\qquad S_{33}=\frac{d(w^{-1})}{d(w)},\\ 
&&S_{11}S_{22}-S_{12}^2=\frac{d^*({w^*}^{-1})}{d(w)},\qquad 
S_{11}S_{33}-S_{13}^2=\frac{d^* (-{w^*}^{-1})}{d(w)},\qquad 
S_{22}S_{33}-S_{23}^2=\frac{d(-w)}{d(w)},\nonumber 
\end{eqnarray} 
where the subscripts in the matrix elements $S_{ij}$ denote 
the channels: $i, j = $1--$\pi\pi$,~2--$K\overline{K}$,~3--$\eta\eta$. 
The $d(w)$ function for the resonance part in these relations is 
\begin{equation}\label{dw_res} 
d_{res}(w)=w^{-\frac{M}{2}}\prod_{r=1}^{M}(w+w_{r}^*) 
\end{equation} 
with $M$ a number of resonance zeros. For the background part $S_B$, 
the $d$ function has the form 
\begin{equation} 
d_B=\mbox{exp}[-i\sum_{n=1}^{3}\frac{\sqrt{s-s_n}}{2m_n} 
(\alpha_n+i\beta_n)]
\label{dbackground} 
\end{equation} 
where 
\begin{eqnarray} 
&& 
\alpha_n=a_{n1}+a_{n\sigma}\frac{s-s_\sigma}{s_\sigma}\theta(s-s_\sigma)+ 
a_{nv}\frac{s-s_v}{s_v}\theta(s-s_v),\\ \nonumber 
&& 
\beta_n=b_{n1}+b_{n\sigma}\frac{s-s_\sigma}{s_\sigma}\theta(s-s_\sigma)+ 
b_{nv}\frac{s-s_v}{s_v}\theta(s-s_v)\nonumber 
\end{eqnarray} 
with $s_\sigma$ the $\sigma\sigma$ threshold and $s_v$ the effective threshold 
due to the opening of many channels in the energy region around 1.5~GeV (e.g., 
$\eta\eta^{\prime},~\rho\rho,~\omega\omega$). These thresholds are determined 
in the analysis. 
 
The expressions (\ref{LeCouteur-Newton}) and (\ref{dw_res}) provide 
the simplest possible parametrization of the resonance part of the $S$ matrix 
for a given number and type of resonances on the uniformization $w$ plane 
keeping unitarity and analyticity. The free parameters (the zeros $w_r$) 
of the $S_{res}$ with a particular number and type of resonances, are fixed 
by fitting to the experimental data. The scenario with the smallest $\chi^2$ 
is chosen as the most probable hypothesis on the condition of the given data set. 
Note that this is an opposite approach to that, e.g., in Refs.~\cite{Caprini08,CCL2006} 
where the $S$ matrix is constructed in the physical region and then it is 
analytically continued to the Riemann surface to find poles, for example 
the $\sigma$-meson pole. In our approach, positions of the poles are obtained 
directly from the fitting. An optimal number of poles is the minimal number 
which guarantees a satisfactory description of the data and which contains 
only the poles significantly improving the fit.  
The poles are introduced according to the formulas for analytic continuations 
of the $S$-matrix elements to all sheets \cite{SBL-prd12,KMS-96}. 
 
The background part $S_{bgr}$ is constructed in the physical region to mimic 
an influence of the other singularities not included explicitly 
in the resonance part $S_{res}$. 
The simple form in Eq.(\ref{dbackground}) includes a response to the opening 
of the channels whose threshold branch points are not taken into account 
explicitly in the uniformizing variable. Values of the fitted parameters 
in Eq.(\ref{dbackground}) ($a$ and $b$) indicate a relative importance of 
these branch points, e.g. a negative or large value of some background 
parameter could suggests that the corresponding branch point should be 
explicitely allowed for in the uniformizing variable. 
Therefore, in choosing the best variant we also require that the background  
contribution is negligible, i.e., the background parameters are small. 
 
In our previous analysis of data on $\pi\pi\to\pi\pi,K\overline{K}$ we took  
into account the left-hand branch point at $s=0$ in the uniformizing  
variable in addition to the $\pi\pi$- and 
$K\overline{K}$-threshold branch points \cite{SBKLN_1206_3438}. In the analysis  
of $\pi\pi\to\pi\pi,K\overline{K},\eta\eta$ we allowed rather for  
the $\eta\eta$-threshold branch point \cite{SBL-prd12}. In the presented 
more elaborate three-channel analysis, unlike in Ref. \cite{SBL-prd12},  
we follow more consistently the spirit of the model-independent description  
obtaining practically zero background of the $\pi\pi$ scattering.   
 
\section{Analysis of the data on isoscalar $S$-wave processes 
$\pi\pi\to\pi\pi,K\overline{K},\eta\eta$ and on decays 
$J/\psi\to\phi\pi\pi,\phi K\overline{K}$} 
 
In the combined analysis of data on the isoscalar $S$ waves of processes 
$~\pi\pi\to\pi\pi,K\overline{K},\eta\eta~$ \cite{Hya73,expd1,expd5,expd6,expd2,expd3} 
we added data on decays $J/\psi\to\phi\pi\pi,\phi K\overline{K}$ from the Mark III 
\cite{MarkIII}, DM2 \cite{DM2} and BES II \cite{BES} Collaborations. 
Formalism for calculating the di-meson mass distributions 
of these decays can be found in Refs.~\cite{MP-prd93,Zou-Bugg-prd94}. 
In this approach the pairs of the pseudoscalar mesons in the final 
states are assumed to have $I=J=0$ and they undergo strong interactions 
whereas the $\phi$ meson behaves as a spectator. 
The amplitudes for $J/\psi\to\phi\pi\pi,\phi K\overline{K}$ decays are 
related with the scattering amplitudes $T_{ij}$ $i,j=1-\pi\pi,2-K\overline{K}$ 
as follows: 
\begin{equation} \label{J->11} 
F(J/\psi\to\phi\pi\pi)=\sqrt{2/3}~[c_1(s)T_{11}+c_2(s)T_{21}], 
\end{equation} 
\begin{equation}\label{J->22} 
F(J/\psi\to\phi K\overline{K})=\sqrt{1/2}~[c_1(s)T_{12}+c_2(s)T_{22}], 
\end{equation} 
where~ $c_1=\gamma_{10}+\gamma_{11}s$ and 
$c_2=\kappa_2/(s-\lambda_2)+\gamma_{20}+\gamma_{21}s$ are functions 
of the couplings of the $J/\psi$ to channels 1 and 2; $\kappa_2$, $\lambda_2$, 
$\gamma_{i0}$ and $\gamma_{i1}$ are free parameters. 
The pole term in $c_2$ approximates possible $\phi K$ states which are not  
forbidden by the OZI rules considering quark diagrams of these processes. 
Obviously this pole should be situated on the real $s$-axis below  
the $\pi\pi$ threshold. 
This is an effective inclusion of the effect of so-called ``crossed channel  
final-state interactions'' in $J/\psi\to\phi K\overline{K}$, 
which was studied largely, e.g., in Ref.~\cite{Guo}. 
The di-meson mass distributions are given as 
\begin{equation} 
N|F|^{2}\sqrt{(s-s_i)\bigl(m_\psi^{2}-(\sqrt{s}-m_\phi)^{2}\bigr)\bigl(m_\psi^2- 
(\sqrt{s}+m_\phi)^2\bigr)} 
\end{equation} 
where $N$ is a normalization constant to data of the experiments determined 
in the analysis: 0.7512, 0.2783, and 5.699 for the Mark III, DM2, and BES II 
data, respectively. 
Parameters of the $c_i$ functions, obtained in the analysis, are 
$\kappa_2=0.0843\pm0.0298$,  $\lambda_2=0.0385\pm0.0251$,  
$\gamma_{10}= 1.1826\pm0.1430$, $\gamma_{11}= 1.2798\pm0.1633$, 
$\gamma_{20}=-1.9393\pm0.1703$, and $\gamma_{21}=-0.9808\pm0.1532$. 
The scattering amplitudes $T_{ij}$ are related to the $S$ matrix as 
\begin{equation} 
S_{ij} = \delta_{ij} + 2\,i\,\sqrt{\rho_i\rho_j}\;T_{ij}\,, 
\end{equation} 
where $\rho_j = \sqrt{1-4m_j^2/s}$. 
 
In the analysis we supposed that in the 1500-MeV region there are two 
resonances: the narrow $f_0(1500)$ and wide $f_0^\prime(1500)$. 
The $f_0 (500)$ state is described by the cluster of type ({\bf a}), 
$f_0(1500)$ by type ({\bf c}), and $f_0^\prime(1500)$ by type ({\bf g}). 
The $f_0 (980)$ is represented only by the pole on sheet~II and shifted pole 
on sheet~III. However, the representation of the $f_0 (1370)$ and $f_0(1710)$ 
states is not unique. These states can be described by clusters either 
of type ({\bf b}) or type ({\bf c}). Analyzing only the processes 
$\pi\pi\to\pi\pi,K\overline{K},\eta\eta(\eta\eta^\prime)$,  
similarly as it was done in~\cite{SBL-prd12}, it is impossible to prefer 
any of four indicated possibilities. Moreover, it was found that the data 
admit two sets of parameters of $f_0(500)$ with a mass relatively near 
to the $\rho$-meson mass, and with the total widths either 
$\approx 600$ or $\approx 930$~MeV, solutions A and B, respectively, 
like in Ref.~\cite{SBL-prd12}. 
 
In the extended combined analysis, adding the data on decays 
$J/\psi\to\phi\pi\pi,\phi K\overline{K}$, one can prefer 
the scenarios when the $f_0(1370)$ is described by the cluster of type 
({\bf b}) and $f_0 (1710)$ by the cluster of type either ({\bf b}) or ({\bf c}). 
To be specific, in the following we shall discuss the case when 
the $f_0 (1710)$ is represented by the cluster of type ({\bf c}). 
 
It is interesting that the di-pion mass distribution of the 
$J/\psi\to\phi\pi\pi$ decay of the BES II data from the threshold 
to $\approx 0.85$~GeV clearly prefers the solution with the wider 
$f_0(500)$ (solution B). A satisfactory description of all analyzed processes 
is obtained with the total $\chi^2/\mbox{NDF}=407.402/(389-51)\approx1.21$ 
where for the $\pi\pi$ scattering, $\chi^2/\mbox{NDF}\approx1.15$, 
for $\pi\pi\to K\overline{K}$, $\chi^2/\mbox{NDF}\approx1.65$, 
for $\pi\pi\to\eta\eta$, $\chi^2/\mbox{ndp}\approx0.87$, and for decays 
$J/\psi\to\phi(\pi\pi, K\overline{K})$, $\chi^2/\mbox{ndp}\approx1.21$. 
 
\begin{figure}[!thb] 
\begin{center} 
\includegraphics[width=0.35\textwidth, angle=270]{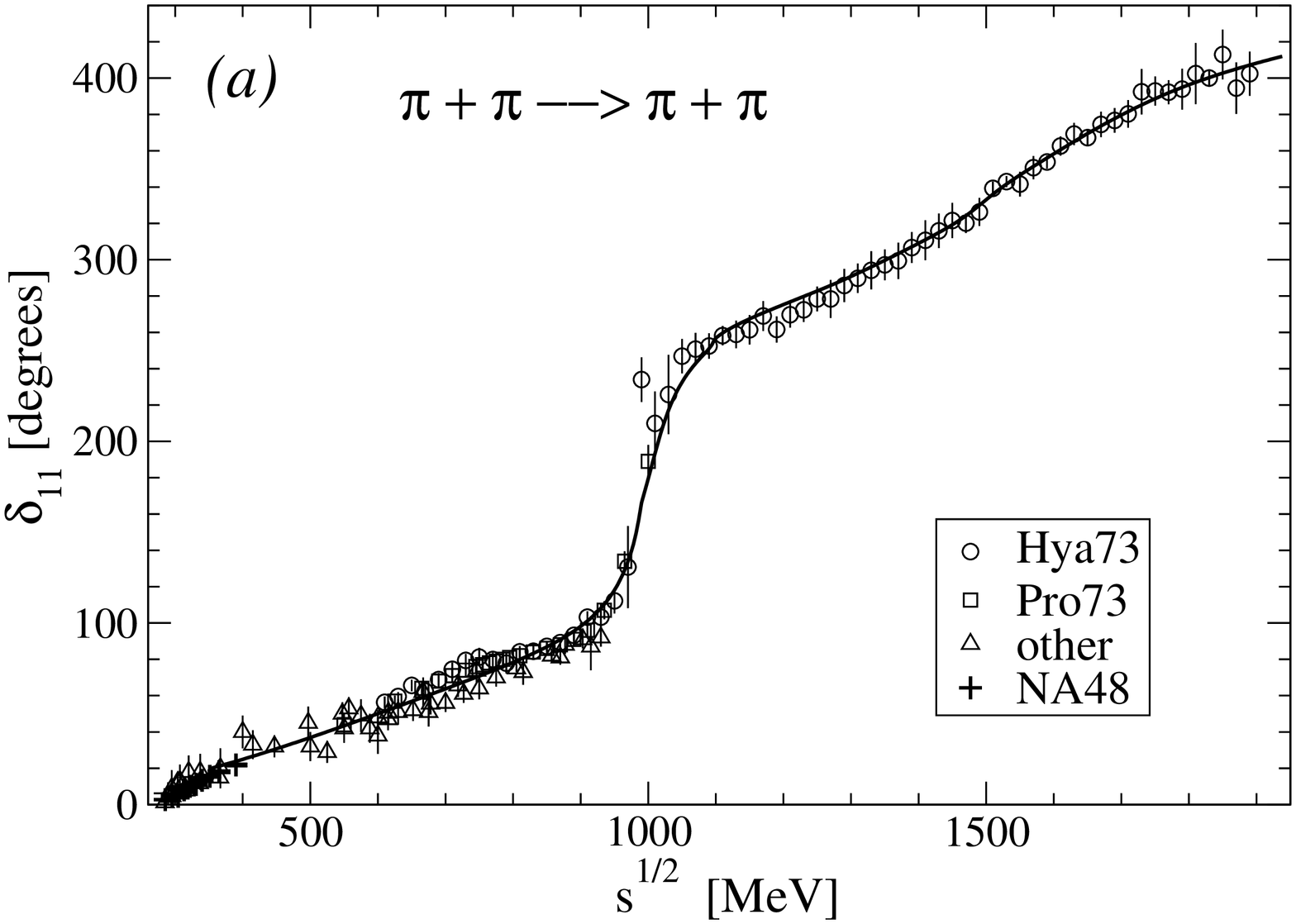} 
\includegraphics[width=0.35\textwidth, angle=270]{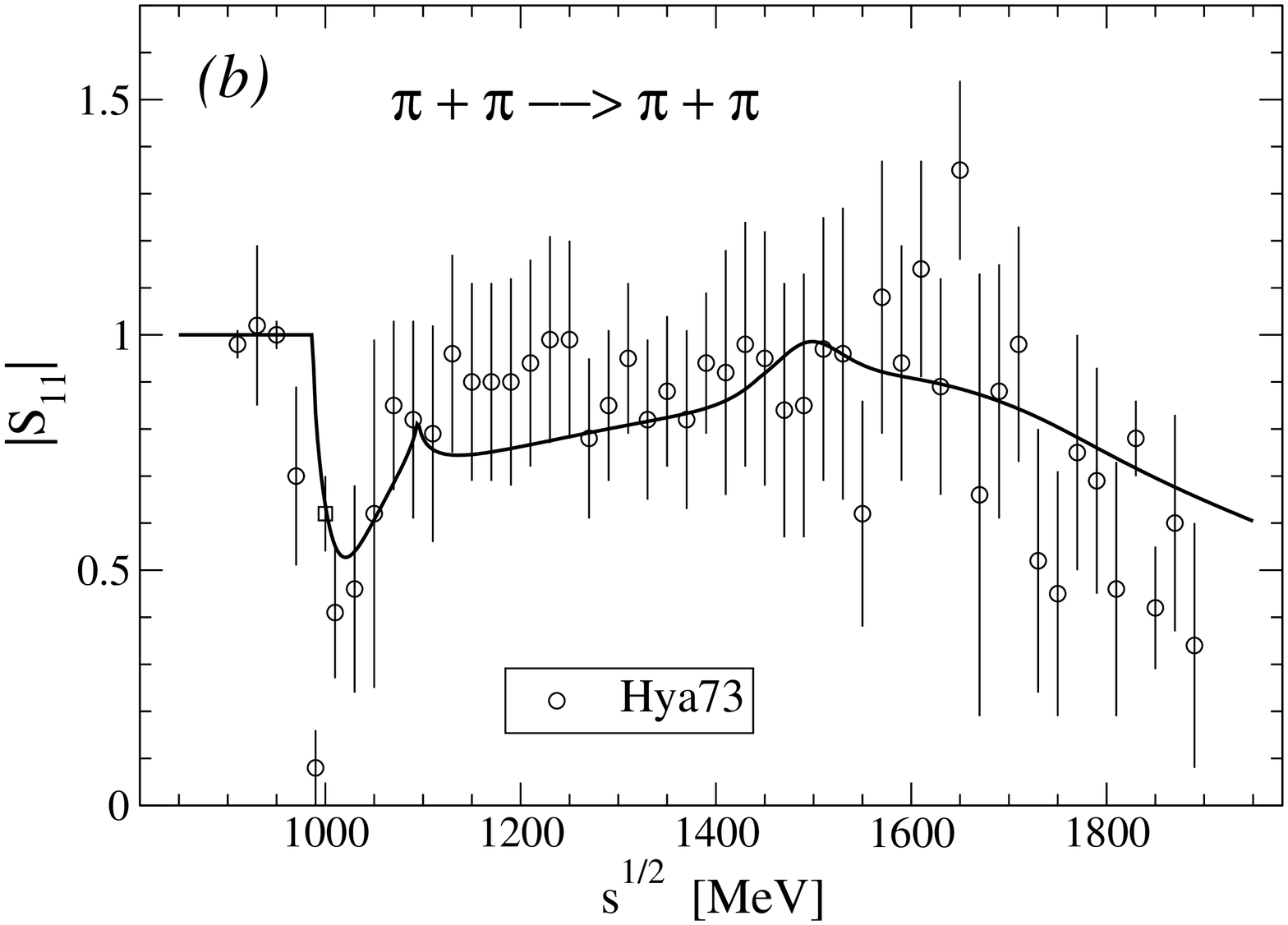} 
\caption{The phase shift ({\it a}) and module ({\it b}) of the 
$\pi\pi$-scattering $S$-wave matrix element. The data are from Ref.~\cite{Hya73} (Hya73), 
\cite{expd5} (Pro73), \cite{expd1} (other), and \cite{NA48} (NA48). \label{fig:pipi}} 
\end{center} 
\end{figure}

\begin{figure}[!thb] 
\begin{center} 
\includegraphics[width=0.35\textwidth, angle=270]{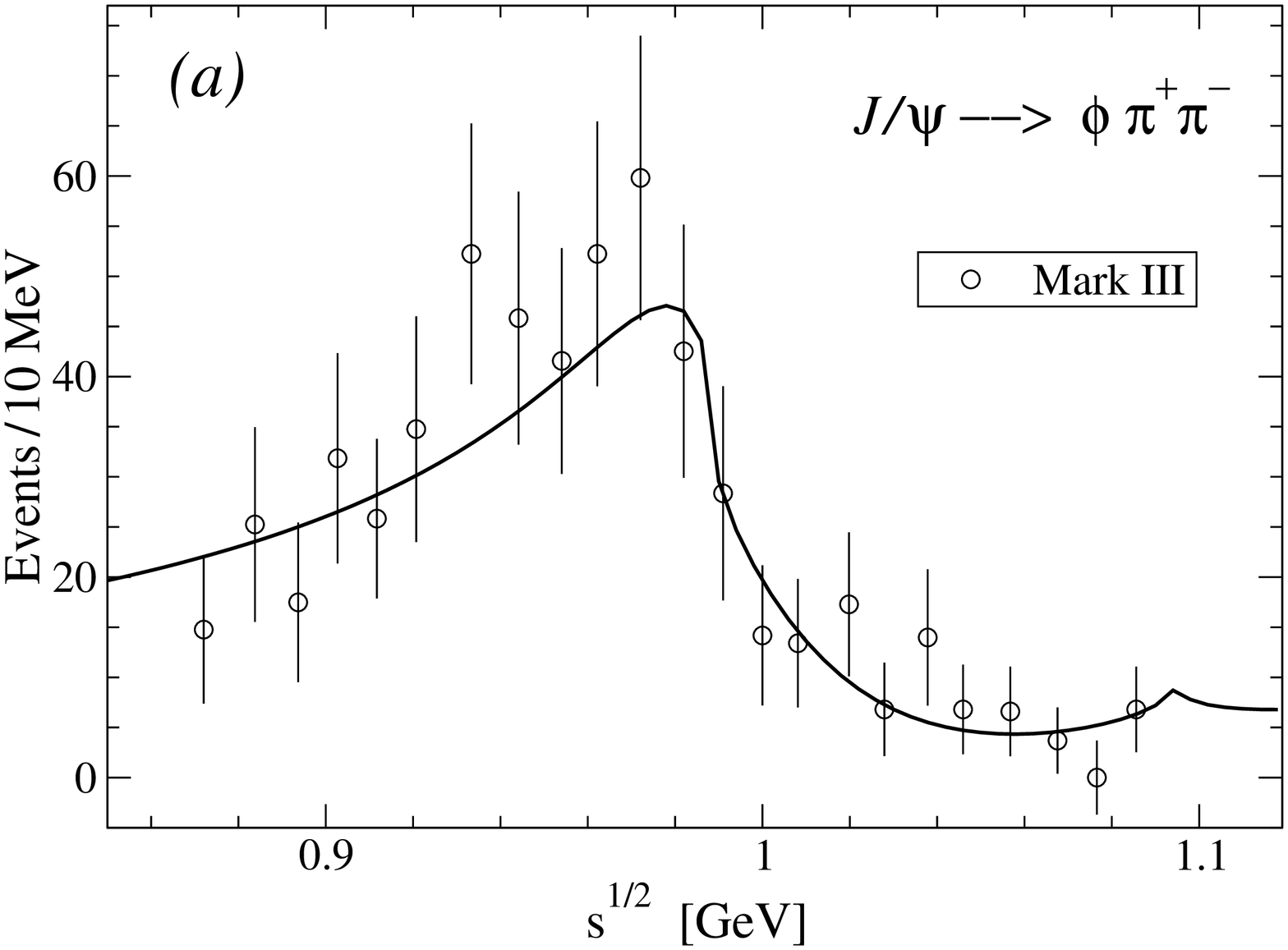} 
\includegraphics[width=0.35\textwidth, angle=270]{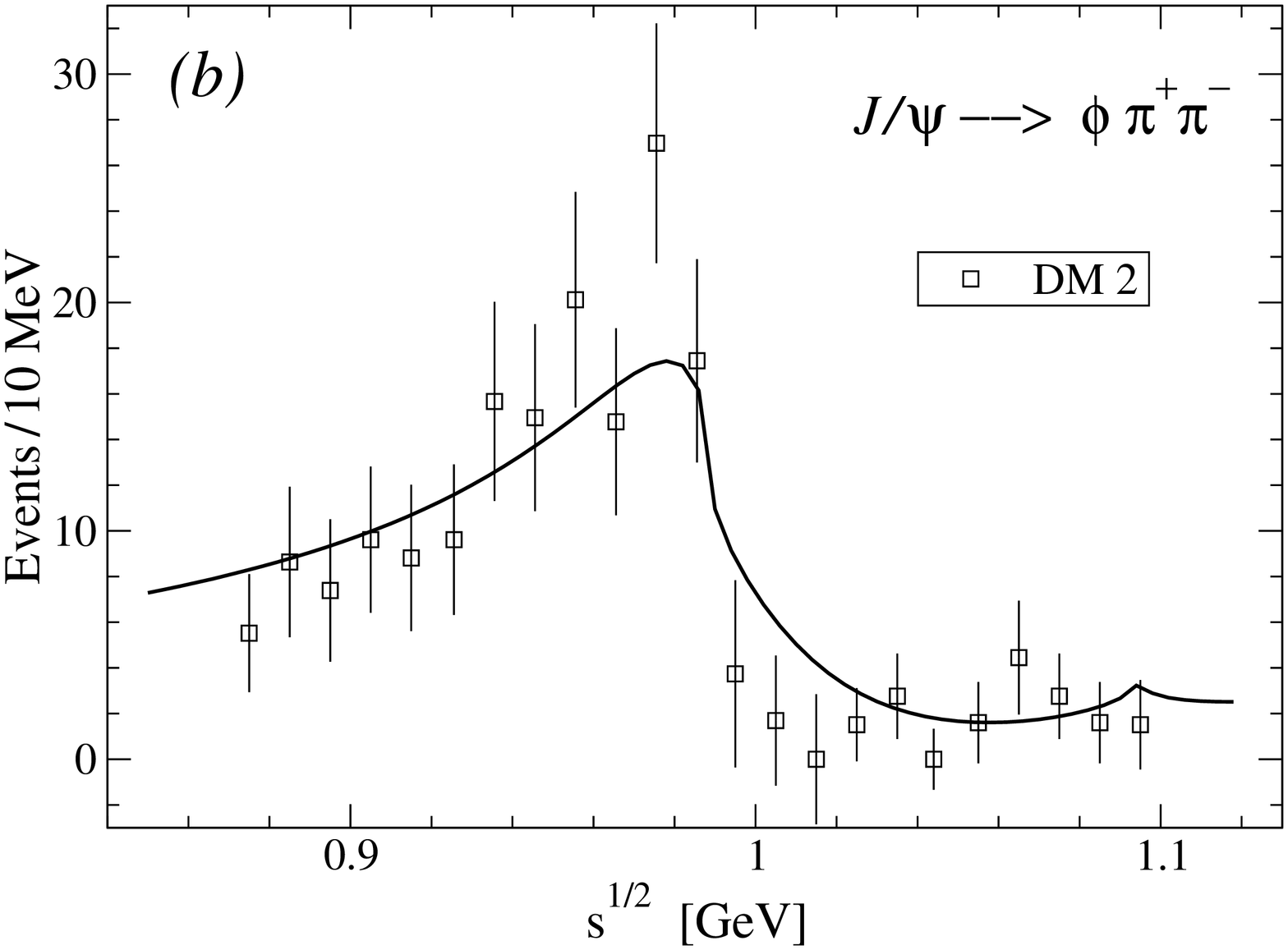} 
\caption{The $\pi^+\pi^-$ invariant mass distributions in 
the $J/\psi\to\phi\pi\pi$ decay. Panel {\it (a)} shows 
the fit to the data of Mark~III and {\it (b)} to DM2. \label{fig:J/psi}} 
\end{center} 
\end{figure}

\begin{figure}[!thb] 
\begin{center} 
\includegraphics[width=0.35\textwidth, angle=270]{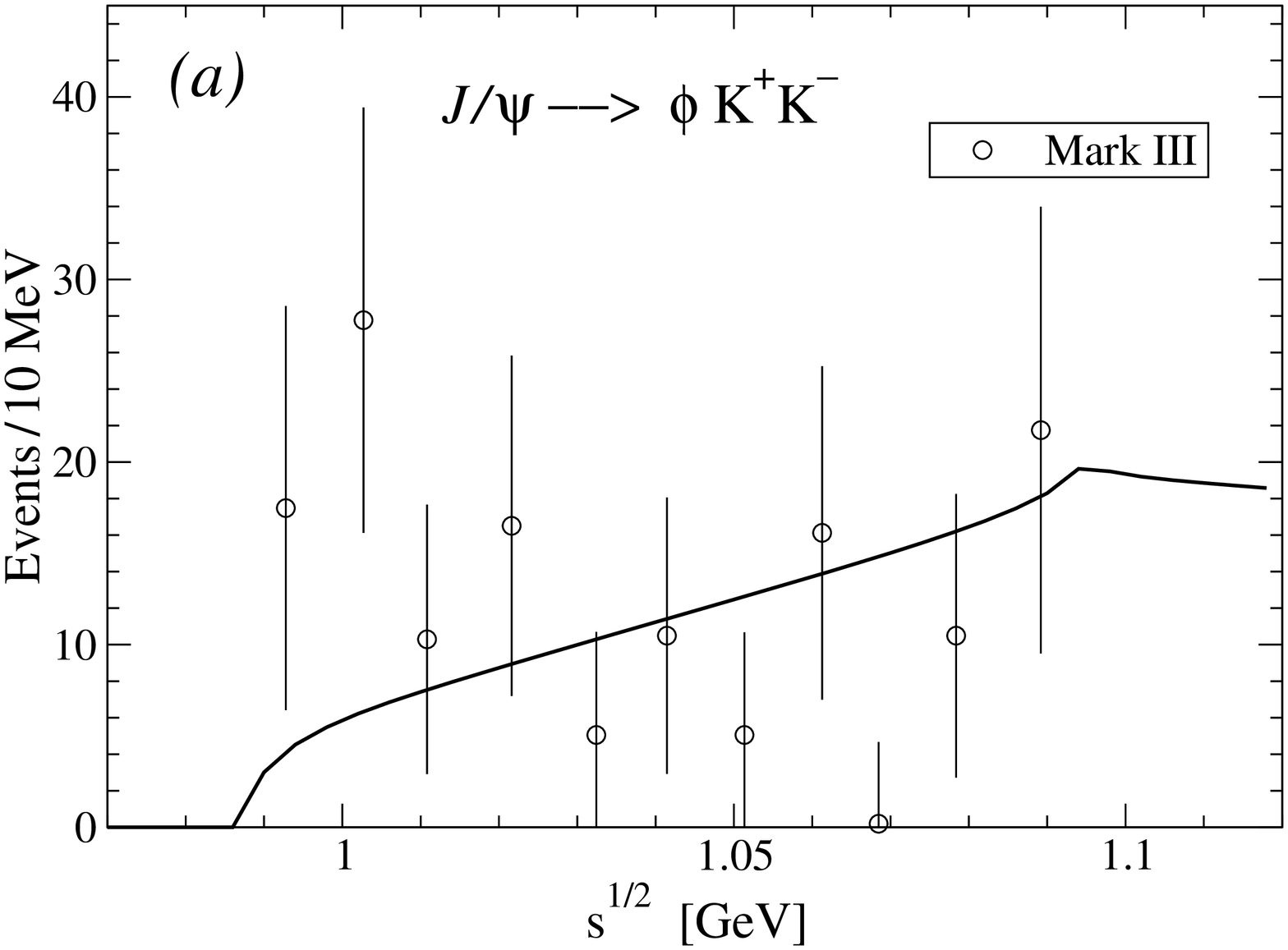} 
\includegraphics[width=0.35\textwidth, angle=270]{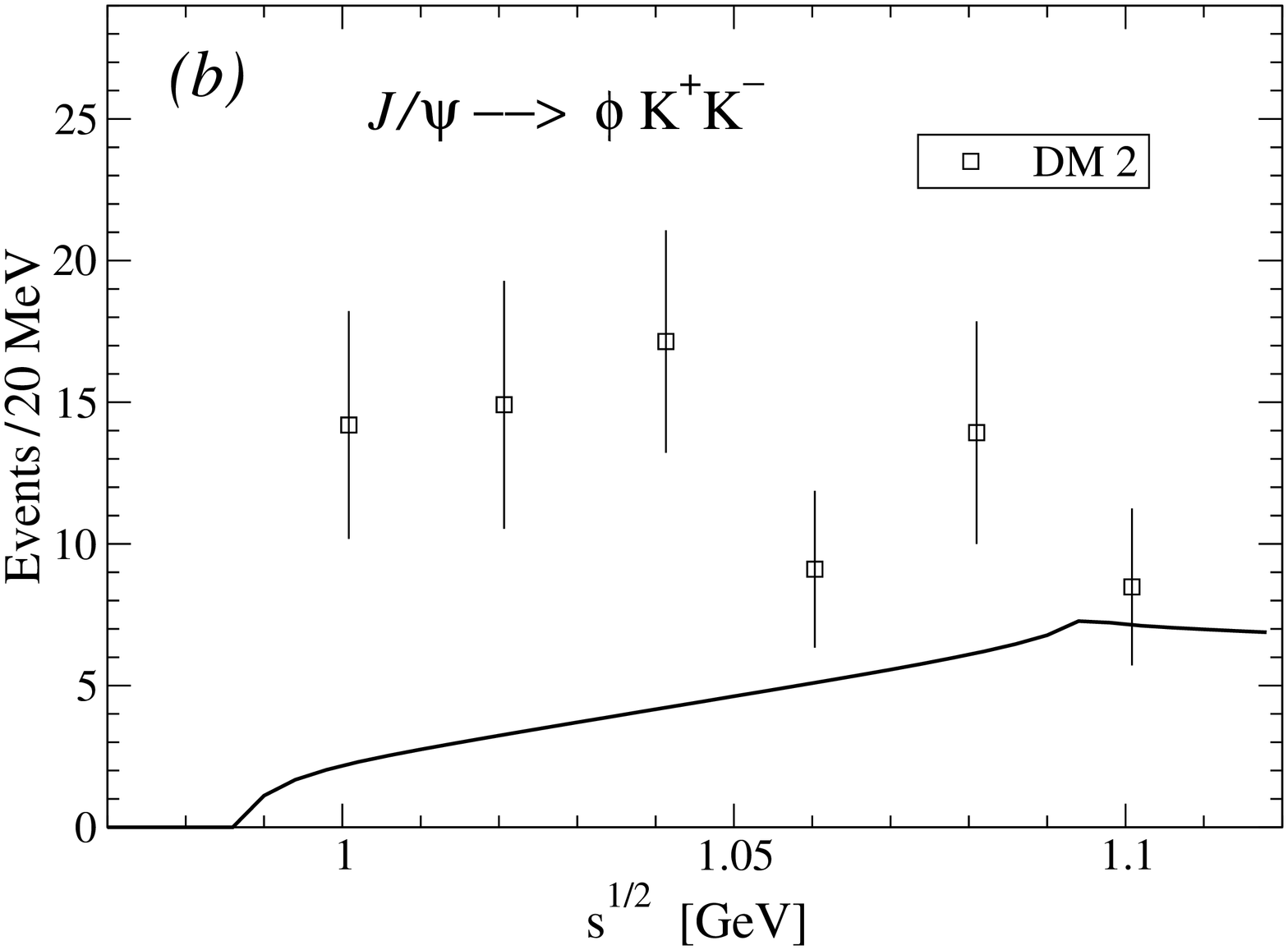} 
\caption{The $K^+K^-$ invariant mass distributions in the 
$J/\psi\to\phi K\overline{K}$ decay. Panel {\it (a)} shows 
the fit to the data of Mark~III and {\it (b)} to DM2. \label{fig:J/psiKK}} 
\end{center} 
\end{figure}

\begin{figure}[!thb] 
\begin{center} 
\includegraphics[width=0.47\textwidth, angle=270]{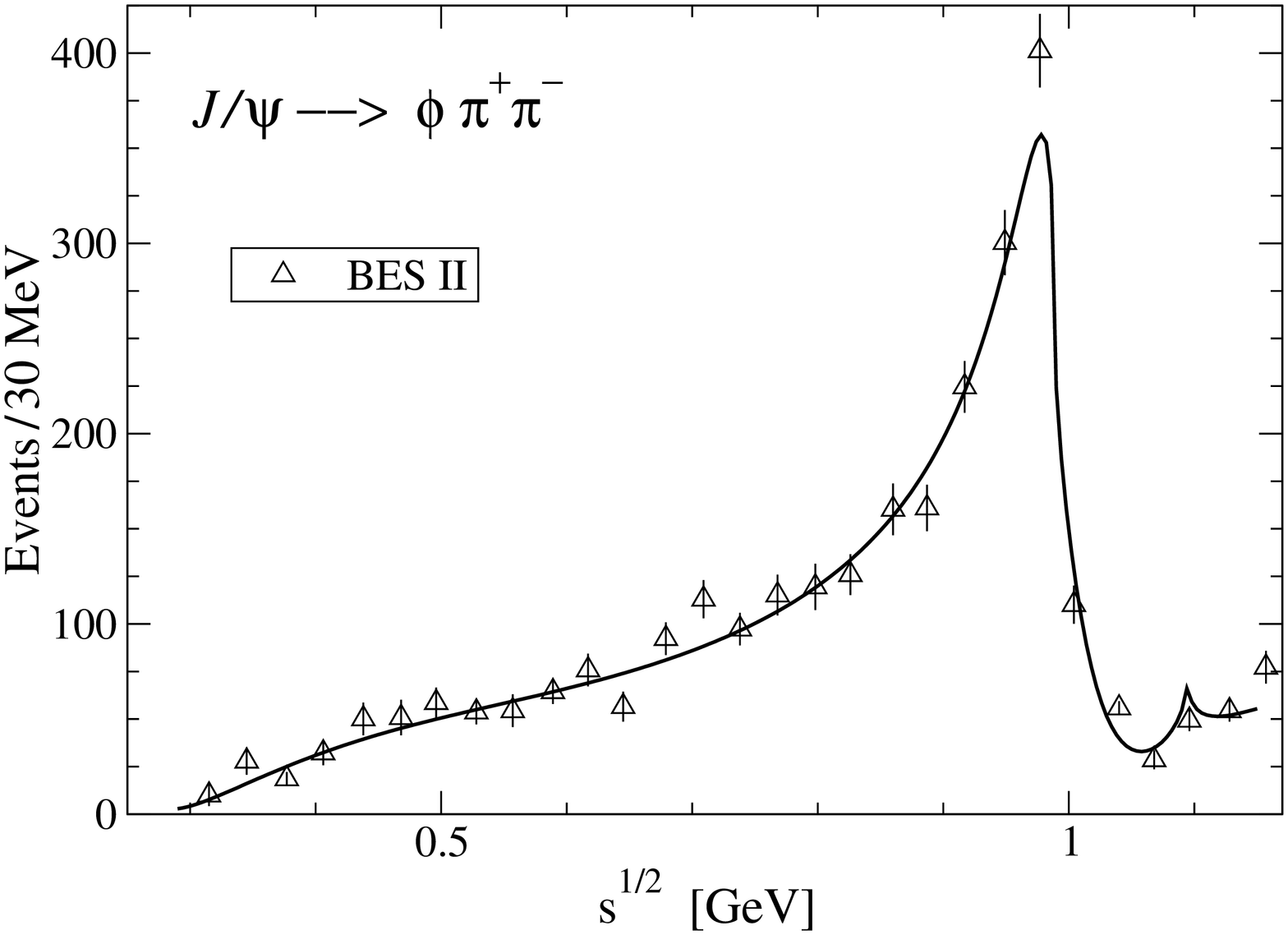} 
\vspace*{-0.2cm}\caption{The $\pi^+\pi^-$ invariant mass 
distribution in the $J/\psi\to\phi\pi\pi$ decay in comparison 
with the data of the BES II Collaboration. \label{fig:J/psi_BES}} 
\end{center} 
\end{figure}

The combined description ($\chi^2$) of processes 
$~\pi\pi\to\pi\pi,K\overline{K},\eta\eta~$ with adding the data on decays 
$J/\psi\to\phi(\pi\pi, K\overline{K})$ is practically the same as in 
Ref.~\cite{SBL-prd12} performed without considering decays of the $J/\psi$ mesons. 
A comparison of the description with the experimental data is shown 
in Figs.~\ref{fig:pipi}-\ref{fig:J/psi_BES}.\vspace{4mm} 
 
%
%
 
%
%
\vspace*{-0.3cm} 
\begin{figure}[!thb] 
\begin{center} 
\includegraphics[width=0.35\textwidth, angle=270]{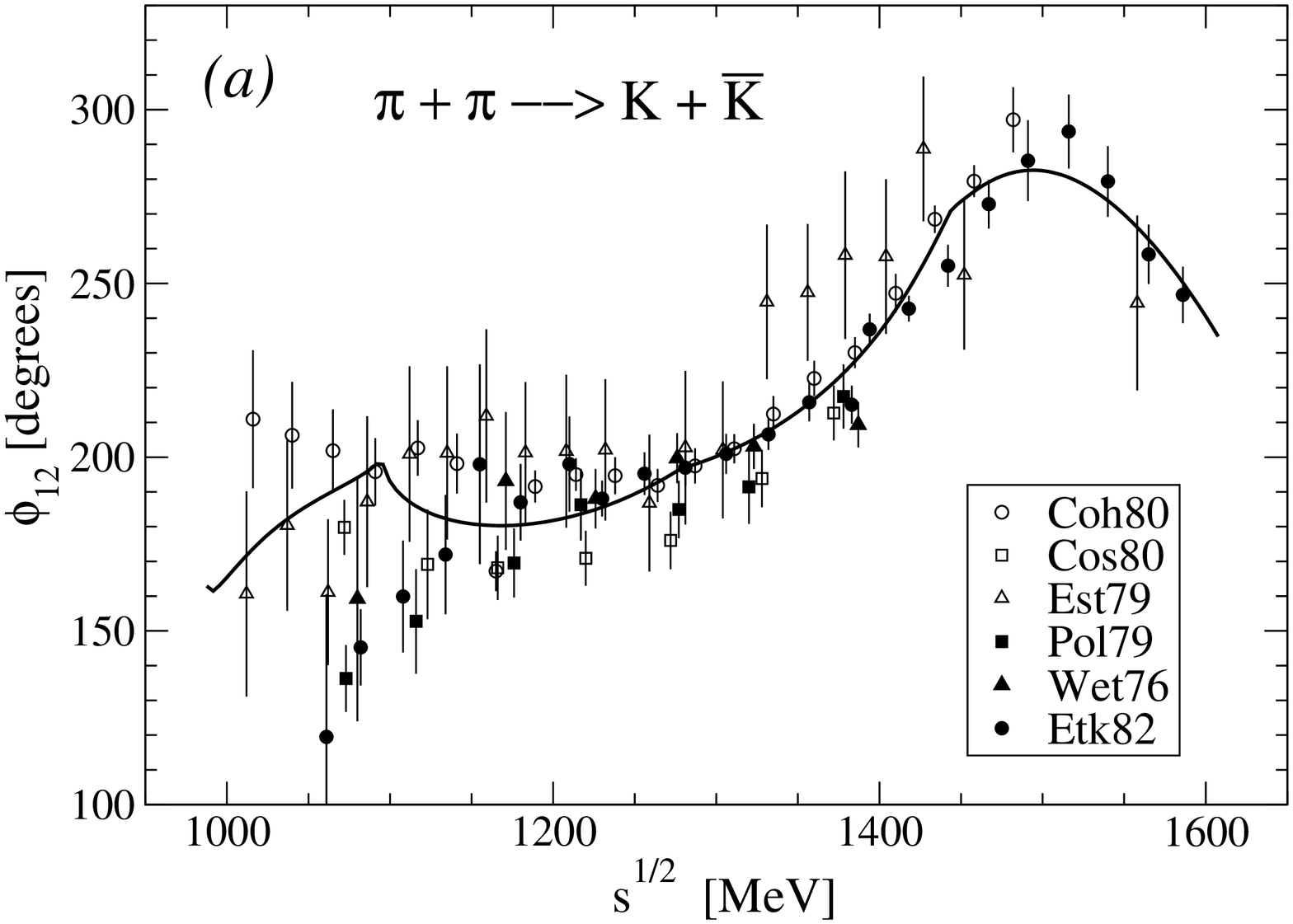} 
\includegraphics[width=0.35\textwidth, angle=270]{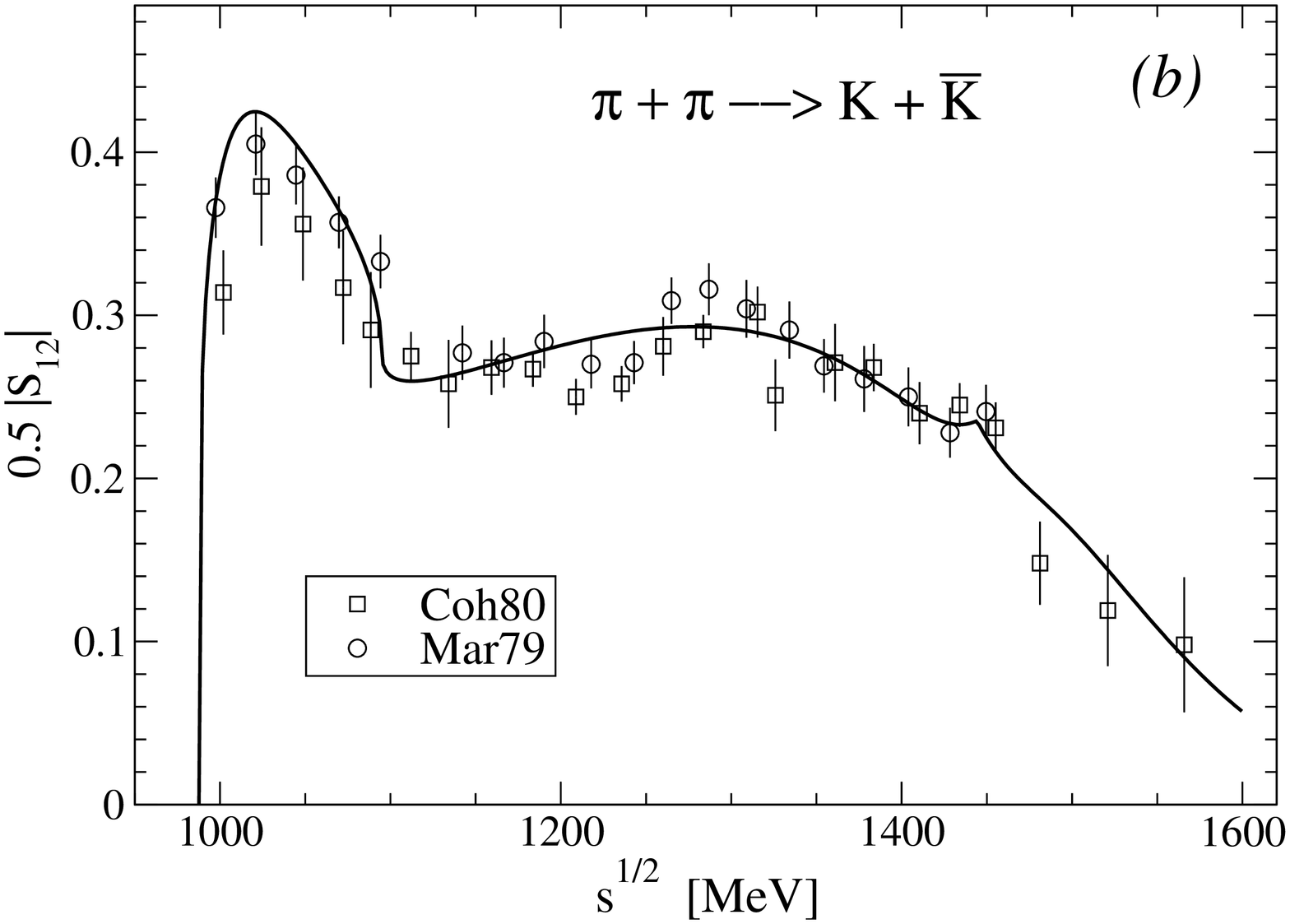} 
\caption{The phase shift {\it (a)} and module {\it (b)} of the 
$\pi\pi\to K\overline{K}$ $S$-wave matrix element. The data are from 
Ref.~\cite{expd2} (Coh80, Cos80, Pol79, Wet76, Etk82, and Mar79) and \cite{expd6} (Est79).} 
\end{center} 
\end{figure} 
 
%
%
\begin{figure}[!thb] 
\begin{center} 
\includegraphics[width=0.37\textwidth, angle=270]{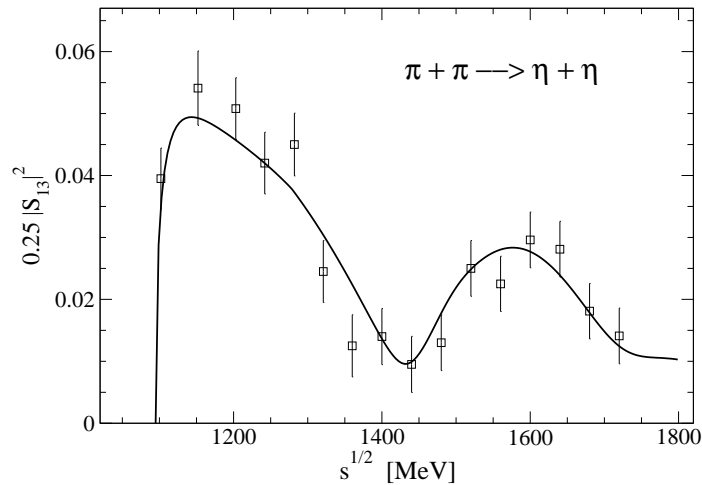} 
\caption{The squared modules of the $\pi\pi\to\eta\eta$ $S$-wave 
matrix element. The data are from \cite{expd3}.} 
\end{center} 
\end{figure} 
 
%
%
 
%
%
 
%
%
 
In Table~\ref{tab:clusters} we show the obtained pole clusters for the 
resonances on the complex-energy plane $\sqrt{s}$. The poles on sheets 
III, V, and VII and VI, corresponding to the $f_0^\prime(1500)$, are of the 
second and third order, respectively (this is an approximation). 
 
%
%
\begin{table}[!htb] 
\caption{The pole clusters for resonances on the $\sqrt{s}$ plane. 
~$\sqrt{s_r}\!=\!{\rm E}_r\!-\!i\Gamma_r/2$ in MeV.} 
\label{tab:clusters} 
\def\arraystretch{1.3} 
\begin{tabular}{|c|c|c|c|c|c|c|c|} 
\hline ${\rm Sheet}$ & {} & $f_0(500)$ & $f_0(980)$ & $f_0(1370)$ & $f_0(1500)$ & $f_0^\prime(1500)$ & $f_0(1710)$ \\ \hline 
II & {${\rm E}_r$} & $514.5\pm12.4$ & $1008.1\pm3.1$\! & {} & {} & $1512.7\pm4.9$ & {} \\ 
{} & {$\Gamma_r/2$} & $465.6\pm5.9$ & $32.0\pm1.5$ & {} & {} & $285.8\pm12.9$ & {} \\ 
\hline III & {${\rm E}_r$} & $544.8\pm17.7$ & $976.2\pm5.8$ & $1387.6\pm24.4$ & {} & $1506.2\pm9.0$ & {} \\{} & {$\Gamma_r/2$} & $465.6\pm5.9$ & $53.0\pm2.6$ & $166.9\pm41.8$ & {} & \!\!\!$127.9\pm10.6$ & {} \\ 
\hline IV & {${\rm E}_r$} & {} & {} & 1387.6$\pm$24.4 & {} & 1512.7$\pm$4.9 & {} \\ 
{} & {$\Gamma_r/2$} & {} & {} & $178.5\pm37.2$ & {} & $216.0\pm17.6$ & {} \\ 
\hline V & {${\rm E}_r$} & {} & {} & 1387.6$\pm$24.4 & $1493.9\pm3.1$ & $1498.9\pm7.2$ & $1732.8\pm43.2$ \\ 
{} & {$\Gamma_r/2$} & {} & {} & $260.9\pm73.7$ & $72.8\pm3.9$ & $142.2\pm6.0$ & $114.8\pm61.5$ \\ 
\hline VI & {${\rm E}_r$} & $566.5\pm29.1$ & {} & 1387.6$\pm$24.4 & $1493.9\pm5.6$ 
& $1511.4\pm4.3$ & 1732.8$\pm$43.2 \\ 
{} & {$\Gamma_r/2$} & $465.6\pm5.9$ & {} & $249.3\pm83.1$ & $58.4\pm2.8$ & $179.1\pm4.0$ & $111.2\pm8.8$ \\ 
\hline VII & {${\rm E}_r$} & $536.2\pm25.5$ & {} & {} & $1493.9\pm5.0$ & $1500.5\pm9.3$ & 1732.8$\pm$43.2 \\ 
{} & {$\Gamma_r/2$} & $465.6\pm5.9$ & {} & {} & $47.8\pm9.3$ & $99.7\pm18.0$ & $55.2\pm38.0$ \\ 
\hline VIII & {${\rm E}_r$} & {} & {} & {} & $1493.9\pm3.2$ & 1512.7$\pm$4.9 & 1732.8$\pm$43.2 \\ 
{} & {$\Gamma_r/2$} & {} & {} & {} & $62.2\pm9.2$ & $299.6\pm14.5$ & $58.8\pm16.4$ \\ 
\hline 
\end{tabular} 
\end{table} 
The pole positions of the $f_0(500)$, $f_0(1370)$, and $f_0(1710)$ have changed with 
respect to Ref.~\cite{SBL-prd12}, especially the first one. 
The pole cluster of $f_0(980)$ practically did not change. 
 
The obtained background parameters are  
$a_{11}=0.0$, $a_{1\sigma}=0.0199\pm0.0052$, $a_{1v}=0.0$, $b_{11}=b_{1\sigma}=0.0$, 
$b_{1v}=0.0338\pm0.0099$, $a_{21}=-2.4649\pm0.0231$, $a_{2\sigma}=-2.3222\pm0.1587$,  
$a_{2v}=-6.611\pm0.5518$, $b_{21}=b_{2\sigma}=0.0$, $b_{2v}=7.073\pm1.287$,  
$b_{31}=0.6421\pm0.0452$, $b_{3\sigma}=0.4851\pm0.1011$, 
$b_{3v}=0$; $s_\sigma=1.6338~{\rm GeV}^2$, $s_v=2.08571~{\rm GeV}^2$.  
The very simple description of the $\pi\pi$-scattering background 
confirms well our assumption in Eq.~(\ref{factorisation}). 
It is important that we have obtained practically zero background of 
the $\pi\pi$ scattering in the scalar-isoscalar channel because a reasonable 
and simple description of the background should be a criterion of correctness 
of the approach. Furthermore, this shows that the consideration of the left-hand 
branch point at $s=0$ in the uniformizing variable solves partly the problem 
of some approaches (see, e.g., \cite{Achasov94}) that the wide-resonance 
parameters are strongly controlled by the nonresonant background. Note also 
that the zero background of the $\pi\pi$ scattering, in addition to the fact 
that $f_0(500)$ is described by the cluster, indicates this state to be the 
resonance (not a dynamically generated state). The point is that after the 
account of the left-hand branch point at $s=0$, the remaining contributions of 
the crossed $u$ and $t$ channels are meson exchanges. 
The elastic background of the $\pi\pi$ scattering is related mainly to 
contributions of the crossed channels. Its zero value means that the exchange 
by the nearest $\rho$ meson is obliterated by the exchange by a particle of near 
mass contributing with the opposite sign (the scalar $f_0(500)$) \cite{SKN-epja}. 
 
Generally, the wide multichannel states are most adequately represented by pole 
clusters located in a specific way on the Rieman surface because 
these clusters give the main model-independent contribution of the resonances \cite{SKN-epja}. 
Positions of the poles are rather stable characteristics for various models, 
whereas masses and widths are very model dependent for the wide resonances. 
However, values of masses are needed, e.g., in the mass relations for 
multiplets. In accordance with the discussion in Sec.~II, we emphasize 
that the masses should be calculated using the poles on sheets II, IV, and 
VIII in dependence on the resonance classification. 
Here we use the formulas 
\begin{equation} 
T^{res}=\sqrt{s}~\Gamma_{el}/(m_{res}^2-s-i\sqrt{s}~\Gamma_{tot})\,, 
\label{propagator} 
\end{equation} 
\begin{equation} 
m_{res}=\sqrt{{\rm E}_r^2+\left(\Gamma_r/2\right)^2}~~~ 
{\rm and}~~~\Gamma_{tot}=\Gamma_r\,, 
\label{hmota} 
\end{equation} 
where $E_r$ and $\Gamma_r/2$ are given in Table~\ref{tab:clusters}. 
The calculated masses and widths for the $f_0$ states are shown in 
Table~\ref{tab:mass_width}.  
Let us note again that the mass of very broad resonances, $f_0(500)$, 
strongly depends on the used formula. 
 
%
%
\begin{table}[!htb] 
\caption{Masses and total widths of the $f_0$ states.}\label{tab:mass_width} 
\def\arraystretch{1.3} 
\begin{tabular}{|c|c|c|c|c|c|c|} 
\hline 
{} & $f_0(500)$ & $f_0(980)$ & $f_0(1370)$ & $f_0(1500)$ & $f_0^\prime(1500)$ & 
$f_0(1710)$\\ \hline 
$m_{res}$[MeV] & 693.9$\pm$10.0 & 1008.1$\pm$3.1 & 1399.0$\pm$24.7 & 1495.2$\pm$3.2 & 
1539.5$\pm$5.4 & 1733.8$\pm$43.2 \\ \hline 
$\Gamma_{tot}$[MeV] & 931.2$\pm$11.8 & 64.0$\pm$3.0 & 357.0$\pm$74.4 & 124.4$\pm$18.4 & 
571.6$\pm$25.8 & 117.6$\pm$32.8 \\ \hline 
\end{tabular} 
\end{table} 
 
\section{Discussion of the results and conclusions} 
 
In the combined model-independent analysis of data on 
~$\pi\pi\to\pi\pi,K\overline{K},\eta\eta$~ in the $I^GJ^{PC}=0^+0^{++}$ 
channel and on $J/\psi\to\phi\pi\pi,\phi K\overline{K}$ from the Mark~III, 
DM2, and BES II Collaborations, an additional confirmation of the $f_0(500)$  
with the pole at $514.5\pm12.4-i(465.6\pm5.9)$ MeV on sheet~II is obtained, 
which can be related with the mass $694\pm 10$~MeV and width 
$931\pm 12$~MeV via Eq.~(\ref{hmota}). 
The real part of the pole is in a good agreement with the results of other 
analyses cited in the PDG tables of 2012: The PDG estimation for the $f_0(500)$ 
pole is $(400\div550)-i(200\div350)$ MeV. 
The obtained imaginary part is, however, larger than the PDG estimation. 
As large values of the imaginary part of the $f_0(500)$ pole appear 
to be inherent in our method of analysis~\cite{SBL-prd12,SBKLN-PRD12}, 
the origin of this interesting result should be understood. 
In Ref.~\cite{SBKLN-PRD12} we showed that a relatively narrow $\sigma$ meson 
consistent with PDG can be obtained in the analysis of one-channel $\pi\pi$ 
scattering data but the inclusion of the $K\bar{K}$ data into the analysis 
makes the width significantly larger. Therefore it seems that the large width 
is tightly connected with the multichannel analysis of data. 
The value of the mass, which gets a significant contribution from the large 
width, Eq.~(\ref{hmota}), agrees well with the prediction by Weinberg in  
Ref.~\cite{Weinberg90}. 
In this work it was shown that even where the chiral symmetry is spontaneously  
broken it can still be used to classify hadron states. Such mended symmetry  
leads to a quartet of particles with definite mass relations and C parity,  
giving the prediction $m_{\sigma}\approx m_\rho$. 
This prediction is also in agreement with a refined analysis using the 
large-$N_c$ consistency conditions between the unitarization and resonance 
saturation which suggests $m_\rho-m_\sigma=O(N_c^{-1})$ \cite{Nieves-Arriola}. 
In addition, in the soft-wall anti-de Sitter/QCD approach~\cite{GLSV_13} -- 
the approach based on gauge/gravity duality --  
the predicted mass of the lowest $f_0$ meson, 721~MeV, practically 
coincides with the value obtained in our work. The above discussion concerns 
solution B, which is prefered by the analysis presented in this paper. The imaginary 
part of the $f_0(500)$ pole in solution A (343 MeV) \cite{SBL-prd12} 
is still in agreement with the PDG estimation. However, solution A is 
inconsistent with data on the $J/\psi\to\phi\pi\pi$ decay from the BES II 
Collaboration: The corresponding curve in Fig.~\ref{fig:J/psi_BES} lies 
considerably below the data from the threshold to about 850 MeV. 
Therefore, solution A is not considered in this paper. 
Anyway the question of too large width of the $f_0(500)$ desires a further 
investigation, estimating the theoretical uncertainties of our approach. 
 
The obtained results for $f_0(980)$, $m_{res}=1008\pm 3$ MeV and 
$\Gamma_{tot}=64\pm 3$~MeV, indicate that the $f_0(980)$ is a non-$q\bar{q}$ 
state, e.g., the $\eta\eta$ bound state because it lies slightly above 
the $K\overline{K}$ threshold and is described by the pole on sheet II 
and by the shifted pole on sheet III without the corresponding (for standard 
clusters) poles on sheets VI and VII. In the PDG tables of 2010 its mass is 
980$\pm$10~MeV. We found in all combined analyses of the multichannel $\pi\pi$ 
scattering the $f_0(980)$  slightly above 1~GeV, as in the dispersion-relations 
analysis only of the $\pi\pi$ scattering \cite{GarciaMKPRE-11}. In the PDG tables 
of 2012, for the mass of $f_0(980)$ an important alteration appeared: Now there 
is given the estimation 990$\pm$20~MeV. 
 
We conclude that the ${f_0}(1370)$ and $f_0 (1710)$ states are dominated 
by the $s{\bar s}$ component in the wave function. The conclusion about 
the ${f_0}(1370)$ agrees with results of the work of the Crystal Barrel 
Collaboration \cite{Amsler} where the ${f_0}(1370)$ is identified as 
the $\eta\eta$ resonance in the $\pi^0\eta\eta$ final state of the ${\bar p}p$ 
annihilation at rest. 
This also explains well why one did not find this state considering 
only the $\pi\pi$ scattering process \cite{Ochs_Mink}. 
The conclusion about the $f_0 (1710)$ is consistent with the experimental 
fact that this state is observed in $\gamma\gamma\to K_SK_S$ \cite{Braccini} 
but it is not observed in $\gamma\gamma\to\pi^+\pi^-$ \cite{Barate}. 
 
In the 1500-MeV region, indeed, there are two states: the $f_0(1500)$ 
($m_{res}\approx1495$~MeV, $\Gamma_{tot}\approx124$~MeV) and the 
$f_0^\prime(1500)$ ($m_{res}\approx1539$~MeV, $\Gamma_{tot}\approx574$~MeV). 
The $f_0^\prime(1500)$ is interpreted as a glueball taking into account 
its biggest width among the enclosing states \cite{Anis97}. As to the large 
width of the glueball, it is worth to indicating Ref.~\cite{Ellis-Lanik}. 
There an effective QCD Lagrangian with the broken scale and chiral symmetry 
is used, 
where a glueball is introduced to theory as a dilaton and its existence is 
related to the breaking of scale symmetry in QCD. The $\pi\pi$ decay width of 
the glueball, estimated using low-energy theorems, is $\Gamma(G \to 
\pi\pi)\approx 0.6\,{\rm GeV}\times(m_G/1\,{\rm GeV})^5$ where $m_G$ is 
the glueball mass. That is, if the glueball with the mass of about 1~GeV exists, 
then its width would be near 600~MeV. Of course, the use of the above formula 
is doubtful above 1~GeV; however, a trend for the glueball to be wide is 
apparently seen. On the other hand, in a two-flavour linear sigma model with 
global chiral symmetry and (axial-)vector mesons as well as an additional 
glueball degree of freedom where the glueball is also introduced as a dilaton 
\cite{Parganlija-glbl}, there arises the rather narrow resonance in the 
1500-MeV region as predominantly a glueball with a subdominant $qq$ component. 
On second thoughts, this result can be considered as preliminary due to using 
a quite rough flavor-symmetry SU($N_f = 3$) in the calculations or, e.g., 
evaluating the $4\pi$ decay, the intermediate state consisting of two 
$f_0(500)$ mesons is not included. In Ref.~\cite{GGLF}, where the 
two-pseudoscalar and two-photon decays of the scalars between 1-2~GeV 
were analyzed in the framework of a chiral Lagrangian and the glueball 
was included as a flavor-blind composite mesonic field, the glueball 
was found to be rather narrow. 
 
Taking into account the discovery of isodoublet $K_0^*(800)$ \cite{PDG-12} 
(see also \cite{SBGL-PPN10}), two lower nonets should correspond to two 
existing isodoublets $K_0^*$. We propose the following sets of the SU(3) 
partners for these states excluding the $f_0(980)$ as the non-$q{\bar q}$ 
state \cite{SBL-prd12}: The lowest nonet consists of the isovector 
$a_0(980)$, the isodoublet $K_0^*(800)$, and $f_0(500)$ and $f_0(1370)$ as 
mixtures of the eighth component of the octet and the SU(3) singlet. The next 
nonet could consist of the isovector $a_0(1450)$, the isodoublet 
$K_0^*(1450)$, and two isoscalars $f_0(1500)$ and $f_0(1710)$. Since this 
assignment removes a number of questions that stood earlier when placing the 
scalar mesons to nonets and does not put forth any new ones, we think this is 
the right direction. An adequate mixing scheme is needed, the search for  
which is complicated by the fact that, in this case, there is also a remainder 
of chiral symmetry which, however, makes it possible to predict correctly, 
{\it e.g.}, the $\sigma$-meson mass \cite{Weinberg90}.

\section*{Acknowledgments} 
 
The authors thank Thomas Gutsche and Mikhail Ivanov for useful 
discussions and interest in this work. 
This work was supported in part by the Grant Program of Plenipotentiary 
of the Slovak Republic at JINR, the Heisenberg-Landau Program, the 
Votruba-Blokhintsev Program for Cooperation of the Czech Republic with JINR, 
the Grant Agency of the Czech Republic (Grant No. P203/12/2126), 
the Bogoliubov-Infeld Program for Cooperation of Poland with JINR, 
the DFG under Contract No. LY 114/2-1. 
The work was also partially supported under the project 2.3684.2011 of 
Tomsk State University. This work has been partly supported by the Polish 
NCN Grant No 2013/09/B/ST2/04382.


\begin{thebibliography}{0} 
 
\bibitem{PDG-12} J.~Beringer {\it et al.} (Particle Data Group), 
                 Phys.\ Rev.\ D {\bf 86}, 010001 (2012). 
 
\bibitem{SBL-prd12} Yu.S.~Surovtsev, P.~Byd\v{z}ovsk\'y, and V.E.~Lyubovitskij, 
                 Phys.\ Rev.\ D {\bf 85}, 036002 (2012). 
 
\bibitem{Bugg07} D.~Bugg, Eur. Phys. J. C {\bf 52}, 55 (2007); arXiv: 0710.4452. 
 
\bibitem{Ochs_Mink} W.~Ochs, AIP Conf. Proc. {\bf 1257}, 252 (2010) 252; 
arXiv:1001.4486v1; P.~Minkowski and W.~Ochs, Eur. Phys. J. C {\bf 9}, 283 (1999); 
hep-ph/0209223; hep-ph/0209225. 
 
\bibitem{Geng_Oset} L.S.~Geng and E.~Oset, Phys. Rev. D {\bf 79}, 074009 (2009). 
 
\bibitem{SBKLN_1206_3438} 
Yu.S.~Surovtsev, P.~Byd\v{z}ovsk\'y, R.~Kami\'nski, V.E.~Lyubovitskij, and M.~Nagy, 
J. Phys. G {\bf 41}, 025006 (2014); arXiv:1206.3438v3[hep-ph]. 
 
\bibitem{MarkIII} W.~Lockman, Proceedings of the Hadron'89 Conference, 
ed. F.~Binon {\it et al.} (Editions Fronti{\` e}res, Gif-sur-Yvette,1989) p. 109. 
 
\bibitem{DM2} A.~Falvard {\it et al.}, Phys.\ Rev.\ D {\bf 38}, 2706  (1988). 
  
\bibitem{BES} M.~Ablikim {\it et al.}, Phys.\ Lett.\ B {\bf 607}, 243 (2005). 
 
\bibitem{Anisovich06} V.V.~Anisovich, Int.~J.~Mod.~Phys. A {\bf 21}, 3615 (2006). 
 
\bibitem{KMS-96} D.~Krupa, V.A.~Meshcheryakov and Yu.S.~Surovtsev, 
Nuovo Cimento A {\bf 109}, 281 (1996). 

\bibitem{SBGKLN-1311.1066} 
Yu.S.~Surovtsev, P.~Byd\v{z}ovsk\'y, T.~Gutsche, R.~Kami\'nski, V.E.~Lyubovitskij, 
and M.~Nagy, arXiv:1311.1066.  

\bibitem{MP-prd93} D.~Morgan and M.R.~Pennington, 
Phys.\ Rev.\ D {\bf 48}, 1185 (1993); {\it ibid.} {\bf 48}, 5422 (1993). 
 
\bibitem{Zou-Bugg-prd94} B.S.~Zou and D.V.~Bugg, 
Phys.\ Rev.\ D {\bf 50}, 591 (1994). 
 
\bibitem{LeCou} K.J.~Le~Couteur, Proc. R. Soc.. A {\bf 256}, 115 (1960); 
R.G.~Newton, J. Math. Phys. {\bf 2}, 188 (1961); 
M.~Kato, Ann. Phys. {\bf 31}, 130 (1965). 
 
\bibitem{Caprini08} I. Caprini, Phys. Rev. D {\bf 77}, 114019 (2008). 
\bibitem{CCL2006} I. Caprini, G. Colangelo, and H. Leutwyler, 
Phys. Rev. Lett. {\bf 96}, 132001 (2006). 
 
\bibitem{Hya73} B. Hyams {\it et al.}, Nucl. Phys. \textbf{B64}, 134 (1973); 
\textbf{100}, 205 (1975). 
 
\bibitem{expd1} A. Zylbersztejn {\it et al.}, Phys. Lett. \textbf{38B}, 457 (1972); 
                P. Sonderegger and P. Bonamy, in Proc. 5th Int. Conference 
                on Elementary Particles, Lund, 1969, 372; 
                J.R. Bensinger {\it et al.}, 
                Phys. Lett. \textbf{36B}, 134 (1971); 
                J.P. Baton {\it et al.}, Phys. Lett. \textbf{33B}, 525 (1970); 
                \textbf{33}, 528 (1970); 
                P. Baillon {\it et al.}, Phys. Lett. \textbf{38B}, 555 (1972); 
                L. Rosselet {\it et al.}, Phys. Rev. D \textbf{15}, 574 (1977); 
                A.A. Kartamyshev {\it et al.}, 
                Pis'ma Zh. Eksp. Theor. Fiz. \textbf{25}, 68 (1977); 
                A.A.~Bel'kov {\it et al.}, 
                Pis'ma Zh. Eksp. Theor. Fiz. \textbf{29}, 652 (1979). 
 
\bibitem{expd5} S.D.~Protopopescu {\it et al.}, Phys. Rev. D \textbf{7}, 1279 (1973). 
 
\bibitem{expd6} P.~Estabrooks and A.D.~Martin, Nucl. Phys. \textbf{B79}, 301 (1974). 
 
\bibitem{expd2} W. Wetzel {\it et al.}, Nucl. Phys. \textbf{B115}, 208 (1976); 
                V.A. Polychronakos {\it et al.}, 
                Phys. Rev. D \textbf{19}, 1317 (1979); 
                P. Estabrooks, Phys. Rev. D \textbf{19}, 2678 (1979); 
                D. Cohen {\it et al.}, Phys. Rev. D \textbf{22}, 2595 (1980); 
                G. Costa {\it et al.}, Nucl. Phys. \textbf{B175}, 402 (1980); 
                A. Etkin {\it et al.}, Phys. Rev. D \textbf{25}, 1786 (1982); 
                A.D.~Martin and E.N.~Ozmutlu, Nucl. Phys. \textbf{B158}, 520 (1979). 
 
\bibitem{NA48} J.R.~Batley {\it et~al.}, Eur. Phys. J. C {\bf 54}, 411 (2008). 
 
\bibitem{expd3} F. Binon {\it et al.}, Nuovo Cimento A \textbf{78}, 313 (1983). 
 
\bibitem{Guo} 
B.~Liu, M.~Buescher, F.-K.~Guo, C.~Hanhart, and U.-G.~Meissner,  
Eur. Phys. J. C {\bf 63}, 93 (2009). 
 
\bibitem{Achasov94} 
N.N.~Achasov and G.N.~Shestakov, Phys. Rev. D {\bf 49}, 5779 (1994). 
 
\bibitem{SKN-epja} 
Yu.S.~Surovtsev, D.~Krupa, and M.~Nagy, Eur.~Phys.~J. A {\bf 15}, 409 (2002); 
Czech. J. Phys. {\bf 56}, 807 (2006). 
 
\bibitem{Weinberg90} S.~Weinberg, Phys. Rev. Lett. {\bf 65}, 1177 (1990). 
 
\bibitem{Nieves-Arriola} J.~Nieves and E.~Ruiz Arriola, 
Phys. Rev. D {\bf 80}, 045023 (2009). 
 
\bibitem{GLSV_13} T.~Gutsche, V.E.~Lyubovitskij, I.~Schmidt, and A.~Vega, 
Phys. Rev. D {\bf 87}, 056001 (2013). 
 
\bibitem{SBKLN-PRD12} 
Yu.S.~Surovtsev, P.~Byd\v{z}ovsk\'y, R.~Kami\'nski, V.E.~Lyubovitskij, M.~Nagy, 
Phys. Rev. D {\bf 86}, 116002 (2012). 
 
\bibitem{GarciaMKPRE-11} R.~Garc{\'i}a-Mart{\'i}n, R.~Kami{\'n}ski, J.R.~Pel{\'a}ez, 
and J.R.~de~Elvira, Phys.\ Rev.\ Lett.\  {\bf 107}, 072001 (2011). 
 
\bibitem{Amsler} C.~Amsler {\it et al.}, Phys. Lett. B {\bf 355}, 425 (1995). 
 
\bibitem{Braccini} S.~Braccini, Frascati Phys. Series {\bf XV}, 53 (1999). 
 
\bibitem{Barate} R.~Barate {\it et al.}, Phys. Lett. B {\bf 472}, 189 (2000). 
 
\bibitem{Anis97} V.V.~Anisovich {\it et al.}, 
Nucl. Phys. Proc. Suppl. A {\bf 56}, 270 (1997). 
 
\bibitem{Ellis-Lanik} J. Ellis and J. L\'anik, 
Phys. Lett. {\bf 150B}, 289 (1985). 
 
\bibitem{Parganlija-glbl} S.~Janowski, D.~Parganlija, F.~Giacosa, and D.~H.~Rischke, 
Phys. Rev. D {\bf 84}, 054007 (2011). 
 
\bibitem{GGLF} F.Giacosa, T.~Gutsche, V.~E.~Lyubovitskij, and 
A.~Faessler, Phys. Rev. D \textbf{72}, 094006 (2005). 
 
\bibitem{SBGL-PPN10} Yu.S.~Surovtsev, T.~Branz, T.~Gutsche, and V.E.~Lyubovitskij, 
Phys. Part. Nucl. {\bf 41}, 990 (2010). 
 
\end{thebibliography}
\end{document}